\documentclass[sigconf]{acmart}
\AtBeginDocument{%
  \providecommand\BibTeX{{%
    \normalfont B\kern-0.5em{\scshape i\kern-0.25em b}\kern-0.8em\TeX}}}

\copyrightyear{2024}
\acmYear{2024}
\setcopyright{rightsretained}
\acmConference[CHI EA '24]{Extended Abstracts of the CHI Conference on Human Factors in Computing Systems}{May 11--16, 2024}{Honolulu, HI, USA}
\acmBooktitle{Extended Abstracts of the CHI Conference on Human Factors in Computing Systems (CHI EA '24), May 11--16, 2024, Honolulu, HI, USA}
\acmDOI{10.1145/3613905.3650921}
\acmISBN{979-8-4007-0331-7/24/05}

\usepackage{microtype}
\microtypecontext{spacing=nonfrench}

\emergencystretch=\maxdimen
\RequirePackage{caption}

\RequirePackage{tabularx}
\usepackage{multirow}
\usepackage{makecell}

\graphicspath{{figures/}}

\begin{document}

\title{Chart What I Say: Exploring Cross-Modality Prompt Alignment in AI-Assisted Chart Authoring}


\author{Nazar Ponochevnyi}
\email{nazar.ponochevnyi@mail.utoronto.ca}
\orcid{0000-0002-5611-0773}
\affiliation{%
  \department{Faculty of Information} 
  \institution{University of Toronto}
  \city{Toronto}
  \state{Ontario}
  \country{Canada}
}
\author{Anastasia Kuzminykh}
\email{anastasia.kuzminykh@utoronto.ca}
\orcid{0000-0002-5941-4641}
\affiliation{%
  \department{Faculty of Information} 
  \department{Department of Computer Science} 
  \institution{University of Toronto}
  \city{Toronto}
  \state{Ontario}
  \country{Canada}
}


\begin{abstract}
Recent chart-authoring systems, such as Amazon Q in QuickSight and Copilot for Power BI, demonstrate an emergent focus on supporting natural language input to share meaningful insights from data through chart creation.
Currently, chart-authoring systems tend to integrate voice input capabilities by relying on speech-to-text transcription, processing spoken and typed input similarly. However, cross-modality input comparisons in other interaction domains suggest that the structure of spoken and typed-in interactions could notably differ, reflecting variations in user expectations based on interface affordances.
Thus, in this work, we compare spoken and typed instructions 
for chart creation. Findings suggest that while both text and voice instructions cover chart elements and element organization, voice descriptions have a variety of command formats, element characteristics, and complex linguistic features. Based on these findings, we developed guidelines for designing voice-based authoring-oriented systems and additional features that can be incorporated into existing text-based systems to support speech modality.

\end{abstract}




\maketitle

\section{Introduction}

Recent chart-authoring systems, such as Amazon Q in QuickSight\footnote{\url{https://aws.amazon.com/q}} and Copilot for Power BI\footnote{\url{https://learn.microsoft.com/en-us/power-bi/create-reports/copilot-introduction}}, demonstrate an emergent focus on supporting natural language input to discover and share meaningful insights from data through chart creation. 
Furthermore, while the majority of these systems are predominantly focused on text input, there is a growing interest in augmenting text with other input modalities, e.g. voice \cite{geminiteam2023gemini,openai2023gpt4}.
Indeed, the integration of voice format provides a lot of promise, for instance, for improving accessibility \cite{Doush01}, facilitating multi-device setups \cite{Tucker01,fraser2020remap}, and offering greater freedom of expression \cite{Badam2017AffordancesOI}. 
Currently, chart-authoring systems tend to integrate voice input capabilities by relying on
speech-to-text transcription \cite{Setlur01, Gao01, Tang01}, processing spoken and typed input similarly. 
The validity of this approach is predicated upon semantic similarities of spoken and typed input in chart-authoring tasks. 
However, cross-modality comparisons of input in other interaction domains, e.g. search \cite{Guy01,melumad2023vocalizing}, suggest that the structure of spoken and typed-in interactions could notably differ, reflecting variations in user expectations based on the interface affordances \cite{Setlur02}. 

In this work, we explore the semantic alignment of free-form spoken (voice modality) and typed (text modality) chart-authoring instructions. 
Through a user study (n=25), we provided participants with diverse scenarios and asked them to instruct an imaginary virtual agent to create a chart to represent the information in each scenario, which allowed us to collect 100 free-form voice instructions for chart creation.
Through a qualitative analysis of the collected instructions, we identified 6 input strategies (e.g., commands and questions) and 22 common elements, categorized into 5 types -- chart elements, element characteristics, element organization, format of command, and linguistic features (Figure \ref{fig:heatmaps}).
We then selected 200 text descriptions from the NLV Corpus \cite{Srinivasan01} and 200 synthetic text descriptions from the nvBench \cite{Luo01} datasets and applied to them the coding scheme developed in the previous stage, assessing its fitness. We found that while both text and voice instructions often cover the basic chart elements and element organization, voice descriptions have a variety of command formats, element characteristics, and complex linguistic features.

Given the well-established importance of the Large Language Model's (LLM) ability to follow instructions \cite{ouyang2022training}, our findings on the differences in how people express chart-authoring instructions in text and voice surface the need for independent processing of spoken and typed prompts.
Based on these findings, we discuss the design implications for voice-based authoring systems and additional features that should be incorporated into text-based systems to support voice modality. 
This work contributes: 
(1) the structure of free-form voice instructions for chart creation and its similarities and differences with equivalent text instructions;
(2) a set of design implications for developing voice-enabled authoring systems for customizable chart creation 
(3) a publicly available dataset of free-form voice descriptions for chart creation to support the development of voice-enabled chart-authoring systems\footnote{\url{https://github.com/CookieLab-UofT/Voice-Chart-Authoring-Instructions-Dataset}}.  

\section{Related Work}
The integration of Natural Language Interfaces (NLIs) into visualization authoring systems provides a number of benefits by 
eliminating the need for extensive knowledge of a particular interface. For example, VisTalk is an authoring tool with a deep learning-based NL interpreter to translate typed NL utterances into editing actions. Evaluation study shows that VisTalk was easy to learn, even without prior visualization experience \cite{Wang01}. Text-to-Viz is a tool that automatically converts typed statements about simple proportion-related statistics to a set of infographics with pre-designed styles. The research demonstrates that Text-to-Viz also minimizes the learning curve \cite{Cui01}.

Text input is considered among the most common modalities for NLIs in chart-authoring (e.g. ADVISor \cite{Liu01}, Pragmatics Principles \cite{Hoque01}, VisTalk \cite{Wang01}, Snowy \cite{Srinivasan02}, Text2Chart \cite{Rashid01}, etc.). However, 
findings from prior studies indicate that text input may present accessibility challenges \cite{Jung01} and users might struggle to formulate direct commands in text \cite{belkin1980anomalous}.
Correspondingly, there has been a growing interest in supporting multimodal interfaces, aiming to overcome the limitations of typed input 
(e.g. Sevi \cite{Tang01}, Eviza \cite{Setlur01}, Analytical Chatbot \cite{Setlur02}, Multi-Modal NLI \cite{Cox01}, DataTone \cite{Gao01}, NL4DV \cite{Arpit01}, etc.).
For example, \citet{Engel01} designed the tangential-sound interface to allow visually impaired users to leverage apparent tools such as Microsoft® Excel®\footnote{\url{https://www.microsoft.com/en-ca/microsoft-365/excel}} and simultaneously generate audio-touch descriptions for charts. \citet{Srinivasan03}, on the other hand, developed the InChorus system that supports pen, touch, and speech input for visual data analysis.
Among modalities to augment text input, voice modality has gained a particular interest. For instance, \citet{Tang01} developed the Sevi system that enables users to perform data analysis tasks using a speech-to-visualization interface. Voice input has the potential to democratize visualization, 
allowing users to simply speak out loud what visualization they would want to see, i.e. provide voice prompts,
potentially improving accessibility \cite{Doush01} and facilitating multi-device setups \cite{Tucker01}. 

Currently, chart-authoring systems tend to integrate voice input capabilities by relying on
speech-to-text transcription \cite{Setlur01, Gao01, Tang01}. This approach implies inherent similarities in processing spoken and typed prompts for AI-assisted chart authoring: the systems learn to support the functionalities expressed in text prompts and voice prompts are expected to trigger the same set of functionalities. However, spoken input is known to have a high degree of freedom of expression \cite{lin2024rambler,mahed2024llms}, limited only by the user's ability to express a query in natural language \cite{Badam2017AffordancesOI}, and comparisons of spoken and written input in other interaction domains \cite{Guy01,melumad2023vocalizing} suggest notable semantic differences. Then, the question becomes whether the required functionalities expressed through text prompts for chart authoring indeed directly correspond to those expressed through voice prompts.

To ensure the ability of a chart-authoring system to follow the instructions expressed through NL prompts, researchers collect visualization specifications into datasets to train and evaluate machine-learning NL interpreters. For instance, 
the nvBench dataset \cite{Luo01} (25,750 pairs of descriptions and visualizations) was synthesized to evaluate models in 105 different domains and 7 types of visualizations. 
To assess the alignment of current tools with people’s expectations for NLIs used in data visualization, previous research investigated the structure of 
NL input. 
\citet{Setlur01} provided participants with five visualizations and collected various utterances people might use when interacting with a chart. The study identified different tasks people attempted to perform, such as searching, filtering, and changing the chart type, and informed the design of the Eviza system. Similarly, \citet{Wang01} conducted a formative study where participants performed the visualization recreation task and found that people usually start with the chart type and encoding, and then follow up with more configurations. Additionally, participants used data labels, properties, and relations to refer to the objects. Based on these findings, the authors developed the VisTalk tool that allows users to edit visualizations using typed natural language input. 
\citet{Srinivasan01} provided ten pairs of tables and charts to 102 participants and asked users to write a text description of the chart to understand how people tend to specify visualization specifications and what types of references they commonly use. The researchers identified four types of phrasing, including commands, queries, questions, and other variations. Additionally, the authors discovered five types of references: attribute, chart type, encoding, aggregation, and design. Lastly, the NLV Corpus dataset of 893 utterances was made publicly available.

Following these established practices, 
we examine the structural and content characteristics of free-form voice instructions for chart creation. Then, by investigating their similarities and differences with equivalent text instructions, we assess the semantic alignment of spoken and written chart-authoring prompts. 

\section{Method}
\subsection{Data Collection}
We first composed three datasets of chart-authoring instruction: in voice (collected through a user study), in text (from NLV Corpus \cite{Srinivasan01}), and synthetic text (from the nvBench \cite{Luo01}).

\subsubsection{Voice Dataset}

\begin{figure*}
    \centering
    \includegraphics[width=16cm,keepaspectratio]{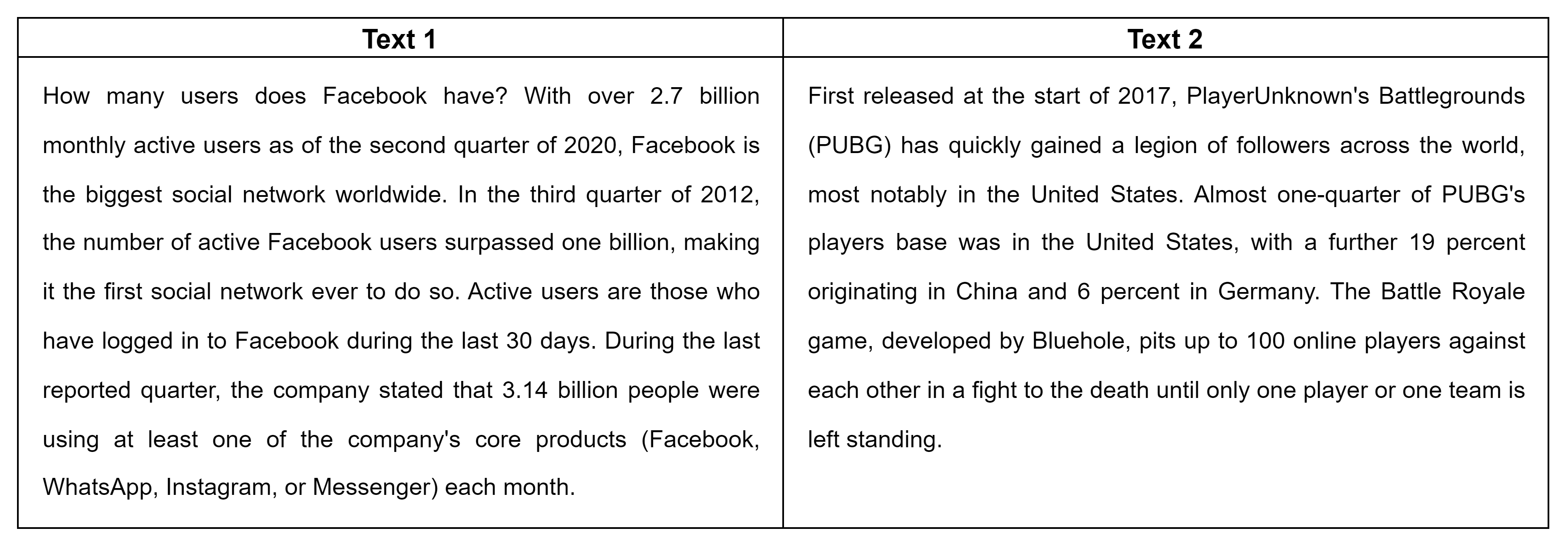}
    \caption{Examples of the text stimuli provided to participants of the user study. From \cite{statista}.}
    \label{fig:stimuli_example}
\end{figure*}

\textbf{User Study Participants.} We recruited participants through email lists targeting students, researchers, and individuals who frequently create charts for work. 
Among 25 participants, all were proficient in English; 
13 self-reported as male, 
12 as female. Participants' self-provided prior experience 
creating charts: None=0, Beginner=7, Intermediate=13, Advanced=5; familiarity with conversational assistants: None=11, Infrequent=3, Regular=8, Expert=3.
\textbf{User Study Stimuli Dataset.} To avoid prompting biases, we 
wanted to avoid 
presenting participants with pre-designed charts and tables of data, and instead, 
provided textual scenario-based stimuli. 
To collect real examples of such texts, we 
searched for materials that often require visualizations to accompany them, such as articles, publications, and reports in economics, geography, statistics, demographics, and news. 
From these texts, we only chose those under 100 words; with content on neutral topics (to avoid potential distress for participants); containing ordinary well-known information, which could be supplemented with visualizations — e.g. numerical values or comparisons of concepts. The final corpus (Fig. \ref{fig:stimuli_example}) included 8 texts from 54 to 100 words (Table \ref{tab:wordcount}). All texts came from the Statista resource \cite{statista}. 

\begin{table}
    \centering
    \begin{tabular}{c|c|c}
        \textbf{Text} & \textbf{Word Count} & \textbf{Collected Instructions} \\
        \midrule
        Text 1 & 96 & 9 \\
        Text 2 & 82 & 9 \\
        Text 3 & 80 & 16 \\
        Text 4 & 65 & 13 \\
        Text 5 & 81 & 14 \\
        Text 6 & 100 & 12 \\
        Text 7 & 54 & 14 \\
        Text 8 & 57 & 13 \\
        \bottomrule
    \end{tabular}
    \caption{Summary of the word count and number of collected instructions for each text stimuli.}
    \label{tab:wordcount}
\end{table}

\textbf{User Study Procedure.} The procedure was approved by the institution's research ethics board and all participants provided informed consent.
The approx. 30-minute moderated study was conducted online via Zoom and audio recorded. Following the instructions, participants were provided with 4 text stimuli, randomly selected from a corpus of 8 texts. 
For each text, participants were asked to imagine a chart to represent the information in it and to describe it verbally as if giving instructions to a virtual assistant on how to create it.
The researchers asked clarifying questions about the participant's instructions throughout the process. 
The study concluded with a semi-structured debriefing interview about the participants' overall experience during the study and their experience working with charts and voice assistants.

Interviews were anonymized, transcribed, and stored on a secure server. 
The final dataset 
included 100 prompts across all texts. 
Given the random assignment of texts to participants, the number of collected instructions per text varied (Table \ref{tab:wordcount}), with an 
av. of 12.5 $\pm$ 2.29. 
We examined various types of charts included in the voice dataset and found 51 bar, 17 line, 8 scatter, 20 pie, and 4 map charts.
Since bar, line, and scatter charts were also present in both NLV Corpus \cite{Srinivasan01} and nvBench \cite{Luo01} datasets, we narrowed our focus to only voice prompts of these charts (n=76).

\begin{figure}[h]
    \centering
    \includegraphics[width=8cm,keepaspectratio]{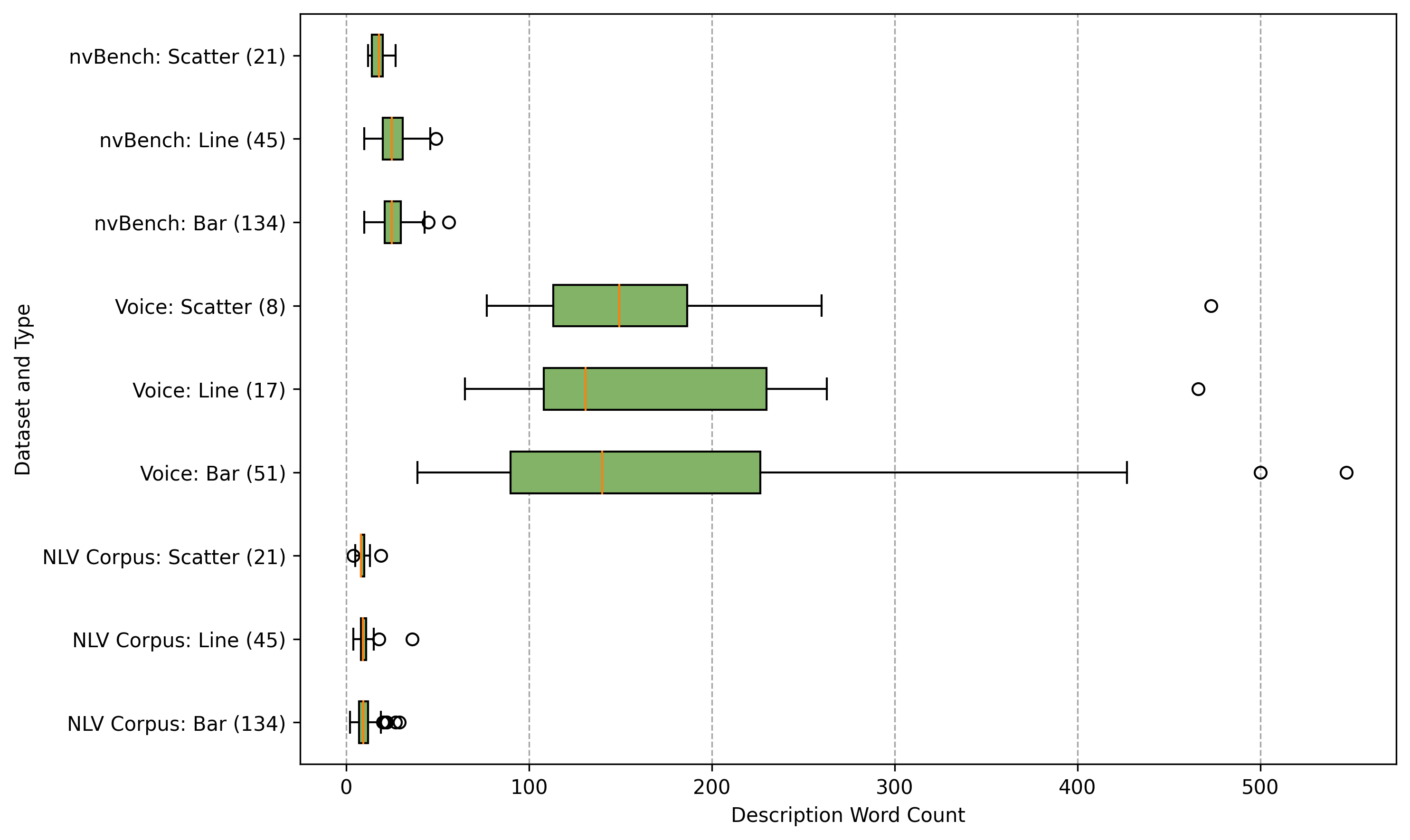}
    \caption{Summary of the word count of the 76 voice, 200 text, and 200 synthetic text instructions.}
    \label{fig:word_count_boxplot}
\end{figure}

\subsubsection{Text Datasets}
To compare voice with the structure of text instructions, we selected only bar, line, and scatter chart types. We randomly selected 
200 text descriptions from the NLV Corpus \cite{Srinivasan01}. Additionally, we wanted to compare it with the synthetic text descriptions, so we randomly chose 200 synthetic text descriptions from the nvBench \cite{Luo01} dataset. We preserved the same ratio between chart types as in our voice dataset. 

\subsection{Data Analysis}




We analyzed voice instructions based on two main aspects: how they are phrased (input strategies) and the information they convey on a particular visualization (elements). The average description word count for the voice dataset -- 175.41 $\pm$ 114.12, the NLV Corpus dataset -- 10.06 $\pm$ 4.58, the nvBench dataset -- 25.19 $\pm$ 7.74 (Fig. \ref{fig:word_count_boxplot}).
Following the qualitative analysis outlined by \citet{braun2006using}, two researchers went through the voice dataset to develop an initial set of open codes, first focusing on instruction elements; 
the codes were then refined in subsequent passes through the data, 
allowing new codes to emerge and for existing codes to evolve as needed. 
Then, researchers collaboratively identified themes by examining the relationships between codes and refined them as necessary, resulting in 22 instruction elements. Finally, the elements were organized into 5 major types (Fig. \ref{fig:heatmaps}).  
To identify input strategies in the description, we applied a top-down approach, starting with 
the coding scheme by \citet{Srinivasan01} and added new codes where needed, identifying 6 input strategies. 
We applied the developed coding scheme to the text-based datasets and assessed its fitness to identify the structural similarities and differences between the voice and text instructions.



\section{Results}

\begin{figure*}
    \centering
    \includegraphics[width=16cm,keepaspectratio]{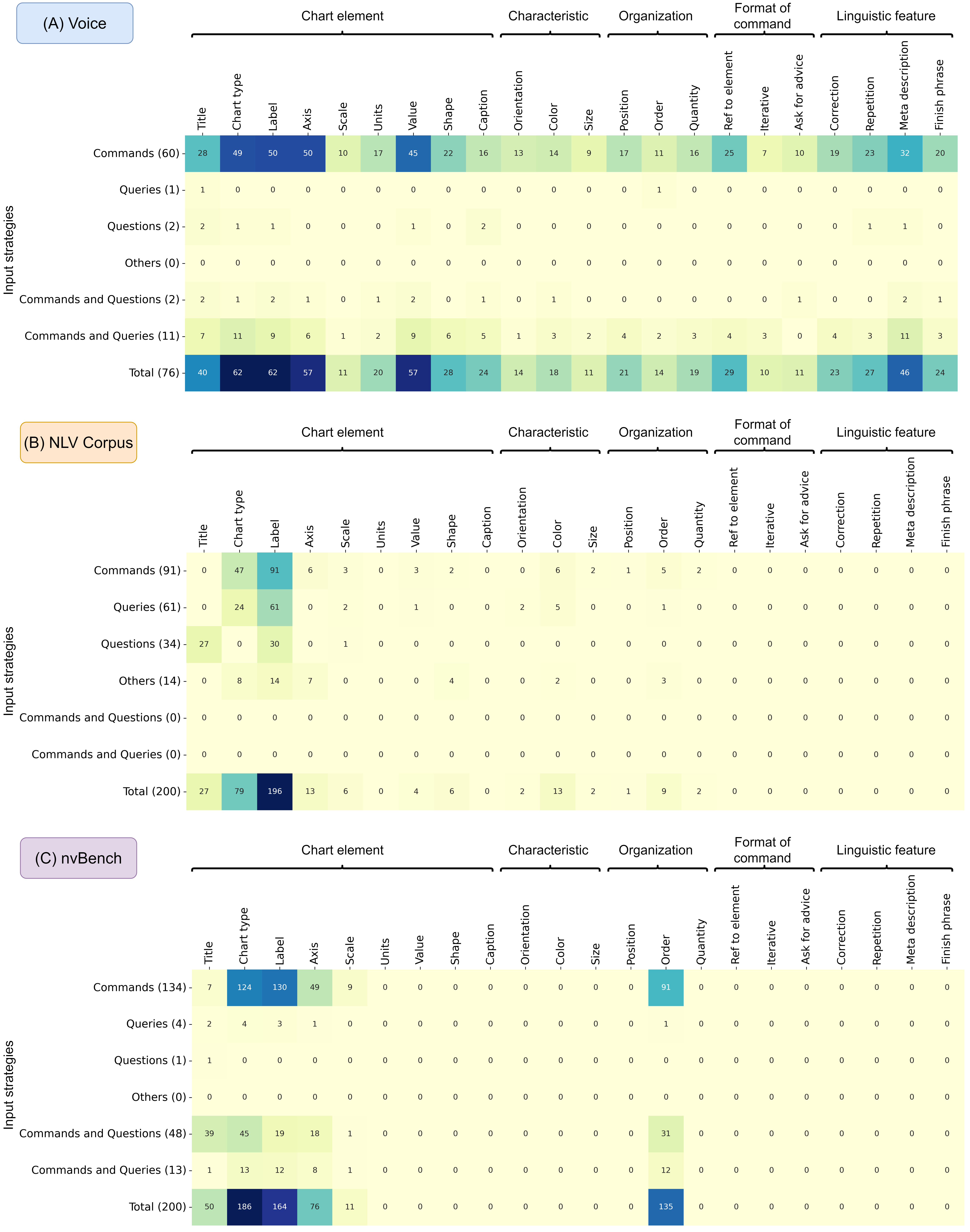}
    \caption{The number of times each element of 5 major types was applied to each input strategy in voice chart-authoring instructions (A), text instructions (B), and synthetic text instructions (C). Color intensity corresponds to data magnitude.}
    \label{fig:heatmaps}
\end{figure*}

Previous research \cite{Srinivasan01} has identified 4 input strategies common in text descriptions: commands (46\% descriptions), queries (30\%), questions (17\%), and others (7\%) (Fig. \ref{fig:heatmaps}.B). Specifically, "commands" are descriptions that sound like instructions, "queries" are descriptions that are similar to search queries, "questions" are descriptions that you can answer with a chart, and "others" are descriptions that do not fit either of these categories, e.g. instructions with special characters that have a programming connotation. In our analysis of voice descriptions, we found that users tend to combine several strategies in one description. We added two additional input strategies: commands and questions (3\%) and commands and queries (14\%) to the existing strategies, such as commands (79\%), queries (1\%), and questions (3\%) (Fig. \ref{fig:heatmaps}.A). Interestingly, synthetic text descriptions also demonstrated some tendencies for combining prompt strategies (Fig. \ref{fig:heatmaps}.C). 

Our qualitative analysis surfaced 5 major types of elements commonly included in voice instructions: chart element (82\% descriptions), element characteristic (24\%), element organization (28\%), format of command (38\%), and linguistic feature (61\%) (Fig. \ref{fig:heatmaps}.A) 
Specifically, chart elements are the main components of the chart, including the title, type, label, axis, scale, units, values, shape, and caption. 
Element characteristics refer to the attributes of the chart elements, such as orientation, color, and size. 
Element organization refers to how participants arrange the chart elements, including their position, order, and quantity. 
Format of command refers to the way in which participants structure their voice commands. 
Participants use iterative commands to perform an action for each element, refer to the chart element created earlier, and ask for advice. 
Linguistic features are the ways in which participants use language to describe the chart elements. This includes correcting previously described elements, repeating descriptions of previously mentioned elements, engaging in self-reflection by providing meta descriptions, and finishing the chart description.
Interestingly, we saw that participants usually did not include a formal finish phrase to signal the completion of their chart description. 
We did not find any significant patterns in the order in which participants provided the various types of voice chart description elements.
We also found that the general structure for pie and map charts in the voice dataset was similar to bar, line, and scatter chart types.
After applying the developed coding scheme to the text-based datasets and assessing its fitness, 
we found that, in text instructions, individuals tend to predominantly specify "chart elements" (98\%) (Fig. \ref{fig:heatmaps}.B): label, chart type, implicit title, and axis. However, they rarely mention "element characteristics" (7\%) and "element organization" (5\%).
Synthetic text descriptions showcased "chart elements" (93\%) and "element organization" (68\%) only (Fig. \ref{fig:heatmaps}.C). Specifically, we found the following elements
: chart type, label, implicit title, axis, scale, and order. 

To summarize, while both text and voice instructions tend to include basic chart elements and element organization, voice prompts have a variety of command formats, element characteristics, and complex linguistic features. 




\section{Discussion and Design Implications}


Previous work on typed NL input \cite{Srinivasan01} discussed that NLIs for data visualization should accommodate natural phrasings (e.g., commands, queries, questions) as part of user input in visualization tools, and recognized the crucial role of the LLM's ability to follow instructions \cite{ouyang2022training}. Wang et al. \cite{Wang01} highlighted that participants use data labels, properties, and relations to refer to the objects. Our study shows that people have higher expectations for voice input in chart creation: while both modalities cover basic elements of chart creation, voice descriptions exhibit greater variety in command formats, element characteristics, and linguistic complexities. 
Voice prompts are generally longer, reflecting a conversational tone and a more verbose communication style, following natural speech patterns more closely than text prompts \cite{Guy01}. We will explore efficient ways of handling lengthy input in future work. In contrast, text instructions are more concise and focused, often employing keywords and a more direct syntax.
Thus, our findings indicate inherent semantic differences between spoken and text-based prompts.

Addressing these differences through a tailored design approach for chart-authoring systems with voice prompting is necessary to accommodate the comprehension and processing of the spoken language's natural flow and complexity and to ensure the system's ability to interpret and execute user instructions accurately.
Voice-based systems should 
be able to effectively parse and interpret unique linguistic structures, 
and recognize and perform on a variety of command formats. 
For example, recognizing commands embedded within longer sentences or interpreting instructions where a user refers to previously mentioned elements.
Voice chart-authoring systems should also support contextual understanding and feedback. This could involve clarifying questions from the system when instructions are ambiguous or advising on chart design when users struggle.

Furthermore, \citet{lester2021power} demonstrated the efficiency of using a single pre-trained model across multiple tasks by only tuning a small set of task-specific parameters. 
A similar approach can be adopted for multimodal chart-authoring systems that accommodate both text and voice inputs, i.e., 
by learning distinct soft prompts for each modality, the system can better recognize and respond to voice and text inputs' unique patterns and interaction structures. 
Our voice dataset can aid in extending existing evaluation protocols, like HELM \cite{liang2023holistic} and SuperGLUE \cite{wang2020superglue}, to more accurately reflect the complexities of spoken chart-authoring prompts. This will enable researchers to design scenarios resembling real-world applications, ensuring a comprehensive assessment of diverse user input styles.

Building on these results, the next step would be to evaluate the accuracy of text and voice input interpretation of chart-authoring systems trained on text and voice data.

\section{Conclusion}
In this work, we demonstrate and categorize semantic differences between spoken and written prompts in the context of AI-assisted chart authoring systems. Building on these findings, we discuss the design implications for integrating functionalities of voice-enabled chart-authoring tools, and provide a unique dataset of voice instructions, serving as a valuable resource for future developments. This research contributes to a broader understanding of multimodal interactions within data visualization contexts, paving the way for more intuitive, efficient, and inclusive authoring systems that leverage the strengths of multimodal prompting.

\begin{acks}
We thank our research participants and anonymous reviewers for their time, effort, and feedback. This work was conducted during the Mitacs Globalink Research Internship at the University of Toronto.
\end{acks}

\bibliographystyle{ACM-Reference-Format}
\bibliography{sample-base}

\end{document}